# Supercooled water confined in a metal-organic framework

Jonas K. H. Fischer[1,3], Pit Sippel[1], Dmytro Denysenko[2†], Peter Lunkenheimer[1*], Dirk Volkmer[2] & Alois Loidl[1]

Within the so-called "no-man's land" between about 150 and 235 K, crystallization of bulk water is inevitable. The glasslike freezing and a liquid-to-liquid transition of water, predicted to occur in this region, can be investigated by confining water in nanometer-sized pores. Here we report the molecular dynamics of water within the pores of a metal-organic framework using dielectric spectroscopy. The detected temperature-dependent dynamics of supercooled water matches that of bulk water as reported outside the borders of the no-man's land. In confinement, a different type of water is formed, nevertheless still undergoing a glass transition with considerable molecular cooperativity. Two different length scales seem to exist in water: A smaller one, of the order of 2 nm, being the cooperativity length scale governing glassy freezing, and a larger one (> 2 nm), characterizing the minimum size of the hydrogen-bonded network needed to create "real" water with its unique dynamic properties.

[1]Experimental Physics V, Center for Electronic Correlations and Magnetism, University of Augsburg, 86135 Augsburg, Germany

[2]Chair of Solid State and Materials Chemistry, Institute of Physics, University of Augsburg, 86135 Augsburg, Germany

[3]Present address: T. Kimura Lab, Department of Advanced Materials Science, University of Tokyo, Japan.

[†]Deceased, June 1st, 2018.

[*] email: peter.lunkenheimer@physik.uni-augsburg.de

The confinement of liquids in spaces of nanometer size has proven a useful tool for the investigation of the glass transition.[1,2,3,4,5,6,7,8,9,10,11] Such experiments can provide valuable information about the increasingly cooperative nature of molecular motion assumed to arise during the gradual transition from a liquid into a glass under cooling.[1,2,3,6,11] Moreover, confinement studies enable to investigate the glassy and supercooled states of liquids that cannot be easily supercooled in their bulk form, avoiding crystallization when encaged in pores holding a limited number of molecules only. One of the most prominent examples in this respect is water,[12,13,14,15,16,17,18,19] a liquid of obvious relevance in numerous fields. Its glassy freezing and a liquid-to-liquid transition were suggested to occur within the so-called "no-man's land" between about 150 and 235 K, where crystallization of its bulk form is inevitable.[20,21,22,23,24,25,26,27] However, these suggestions are highly controversial.[28,29,30,31,32,33,34] Consequently, there are numerous investigations trying to solve this issue by studying the molecular dynamics and glassy freezing of supercooled water in confinement. Many different host materials were employed, e.g., porous glasses and molecular sieves,[13,14,17,35,36,37,38,39,40] clay,[12,41] cement-related gels,[42,43] graphite oxide,[44] and even white bread,[41] to name just a few (see Refs. 18 and 19 for an overview). Unfortunately, many of these materials exhibit highly hydrophilic inner pore surfaces, a considerable pore-size distribution, and/or asymmetric pore geometries, which may hamper the straightforward interpretation of the results.

Different scenarios were proposed to explain the numerous, partly contradicting experimental findings for confined supercooled water, speaking both in favor or against the suggested liquid-to-liquid transition (see Refs. 19, 45, and 46 for examples). Obviously, despite the vast database on confined water, drawing unequivocal conclusions that would lead to a consensus in the community, is difficult. This certainly also is due to the fact that the results considerably depend on the chosen host material.[18,19] Among them, in Ref. 19 water confined in the porous amorphous silica MCM-41, which contains very regular cylindrical pores, was suggested to represent an "ideal model system of confined water". In a very recent work,[45] relating the dynamics of confined water for various host systems to a secondary relaxation, the molecular water dynamics in MCM-41 also was considered to effectively represent the intrinsic bulk-water dynamics.

In the present work, we investigate water confined in a metal-organic framework (MOF), MFU-4*l*-HCOO. Just as for MCM-41, its pores are highly regular and of comparable size. In addition, they are close to spherical, leading to a 3D confinement of the water molecules, in contrast to the highly asymmetric confinement in the channel-like pores of MCM-41.8 In Ref. 47 it was demonstrated that the dimensionality of the pores can influence the properties of confined liquids and, thus, confinement in 3D pores seems preferable for meaningful conclusions concerning the bulk properties. Finally, MFU-4*l*-HCOO exhibits less hydrophilic inner pore surfaces than MCM-41, whose internal surfaces are covered with numerous hydroxyl groups.[48] All these benefits make this MOF an even better-suited host system for the investigation of supercooled water.

In general, MOFs comprise three-dimensional networks with significant porosity which are formed by metal ions or clusters, joined together by bridging organic linkers.[49,50,51,52,53,54,55] Their highly ordered crystal structure ensures well-defined pore sizes. MOFs have outstanding design versatility: For example, by using different organic linkers the pore sizes and the apertures between them can be varied. In two recent reports,[11,56] we have demonstrated that MOFs are well-suited host materials for confinement studies of the glass transition. These works concentrated on the cooperativity length scale of molecular motions in glycerol, a good glass former that can be easily supercooled even in its bulk form. In the present work, we focus on the time and length scales of molecular dynamics when approaching the glass transition of water, a poor glass former whose dynamics when approaching the glass transition cannot be investigated in its bulk form. Water confined in MOFs has been previously investigated numerous times (e.g., 57,58,59), however, mostly not addressing the question of glassy freezing within the no-man's land. To our knowledge, there are only three studies of MOF-confined water using dielectric spectroscopy that touch this topic.[60,61,62] Unfortunately, the characteristic times of water dynamics, detected there, were rather far off those reported for the assumed "ideal" behavior in MCM-41. This fact is probably related to the limited frequency and temperature ranges of these studies and/or to the rather small pore sizes of the used host systems, ranging between about 0.4 and 1.1 nm (compared to about 2 nm for the most common form of MCM-41).

In the present work, we find that water confined in the 1.9 nm pores of MFU-4*l*-HCOO can be well supercooled and exhibits relaxation times close to those found in MCM-41. Thus, this MOF represents a well-suited host for the investigation of the glass transition of confined water. Much clearer than for other confinement hosts, a pronounced non-Arrhenius temperature dependence of the confined-water dynamics is revealed above a crossover temperature $T_{cr} \approx 175$ K. The high data precision, the broad frequency range, the performed evaluation of real and imaginary part including all processes, and the ideal host system available to us allow for a clear proof of this non-Arrhenius behavior. This crossover indicates cooperative molecular motions of supercooled confined water with growing cooperativity length for $T > T_{cr}$. Below $T_{cr}$, the cooperativity length is restricted by the pore size of about 1.9 nm resulting in Arrhenius behavior. Our data provide clear hints at two different length scales associated with glassy cooperativity and an extended hydrogen-bonded network, characterizing the short- and meso-scale molecular order in supercooled bulk water. Such a scenario was only rarely considered previously and never such clear experimental indications were provided.

## Results

**Host Material.** MFU-4*l*-HCOO (where MFU stands for "Metal-Organic Framework Ulm-University") is derived from MFU-4*l*.[63,64] As noted in Ref. 65, the latter is composed of Zn(II) ions and bis(1*H*-1,2,3-triazolo[4,5-*b*], [4',5'-*i*])dibenzo[1,4]dioxin (H$_2$-BTDD) as a ligand. The secondary building unit, [Zn$^o$Zn$^t_4$Cl$_4$(ligand)$_6$], is formed by an octahedrally coordinated zinc ion (Zn$^o$), which is linked to six N-donor atoms of six different ligands, and by four peripheral tetrahedrally coordinated zinc ions (Zn$^t$). Each of the latter are linked to three N-donor atoms of three different ligands. Negatively charged chloride ligands provide charge compensation. In MFU-4*l*-HCOO, the chloride side-ligands are replaced by HCOO-anions[66] (see Fig. 1 for a visualization of its crystal structure).

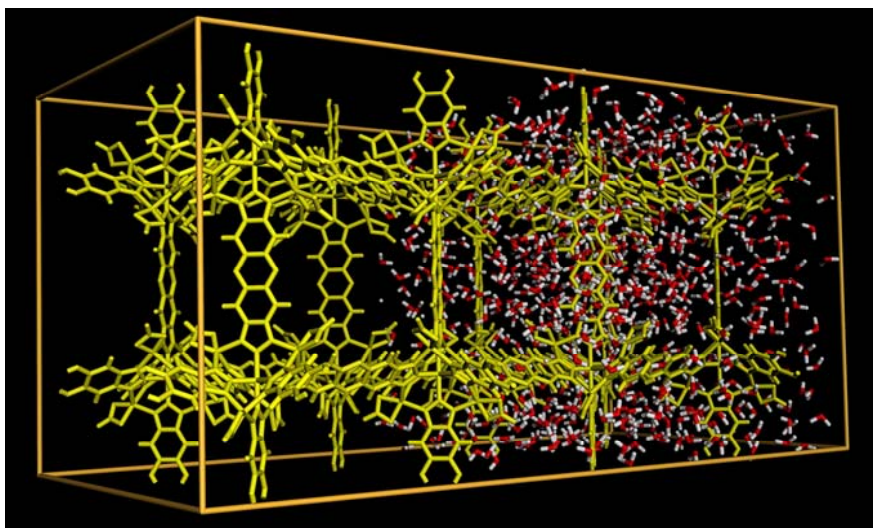

**Fig. 1** Stick model of two adjacent unit cells in MFU-4*l*-HCOO. For the right unit cell, the theoretical maximum filling with water molecules as resulting from simulations is indicated. The model demonstrates the three-dimensional nature of the pores in this material.

The MFU-4*l*-related MOFs comprise two types of pores of different size. MFU-4*l*-HCOO was carefully chosen to achieve a diameter of the bigger pores (1.86 nm) that comes close to the channel diameter of the cylindrical pores in the "ideal" confinement host MCM-41 (2.1 nm for its most common variant). The smaller pores have a diameter of about 1.15 nm. As revealed by Fig. 1, in contrast to MCM-41, the pores have a highly symmetric, three-dimensional shape. Substituting HCOO$^-$ for the Cl$^-$ ligands, pointing inside the pores, reduces the sizes of the apertures between adjacent pores, leading to more effective confinement. The pore diameters of MFU-4*l*-HCOO were derived from the atomic coordinates in its crystalline lattice as reported in Ref. 66, based on X-ray measurements. As these types of MOFs represent rigid covalent networks, the pore sizes should be unaffected by the water filling.

Compared to MCM-41, revealing numerous hydroxyl groups in its inner surfaces,[48] MFU-4*l*-HCOO represents an almost completely hydrophobic network. The only significant interaction of its pore walls with water may arise from the eight negatively-charged formate groups pointing into its pores. Compared to the Cl$^-$ ligands of MFU-4*l*, the more delocalized charge of the HCOO$^-$ ligand should further reduce these interactions. Thus, overall in MFU-4*l*-HCOO the pore-wall interactions of the confined water molecules should be significantly weaker than in other host systems like MCM-41.

To estimate the theoretical maximum filling grade of MFU-4l-HCOO with water molecules, simulations were performed (see Methods section) leading to about 720 H$_2$O molecules per unit cell. The right part of Fig. 1 shows the filling with water molecules as resulting from the simulations.

**Permittivity spectra.** Figure 2 presents spectra of the real and imaginary parts of the dielectric permittivity ($\varepsilon'$ and $\varepsilon''$, respectively), as measured at various temperatures for water in MFU-4*l*-HCOO. As commonly found for confined supercooled liquids, several relaxational processes are detected, signified by a steplike decrease in $\varepsilon'(\nu)$ and peaks or shoulders in $\varepsilon''(\nu)$, which continuously shift to lower frequencies for decreasing temperature. There is no trace of any abrupt qualitative change of the spectra when temperature is decreased that would point to a crystallization of confined water. This is confirmed by differential-scanning-calorimetry measurements between 100 and 200 K, which reveal heat-flow curves without anomalies, making crystallization of any significant fractions of water in MFU-4l-HCOO unlikely (see Supplementary Figure 1).

The found four spectral contributions are numbered 1 - 4 as indicated in Fig. 2. For comparison, we have also measured water-free MFU-4*l*-HCOO with empty pores. As an example, the plusses in Fig. 2 show the results for 189 K, which are featureless and, in case of $\varepsilon''$, close to the instrument resolution. This proves that the processes detected for the MFU-4*l*-HCOO sample with water indeed are related to the water filling. Process 1 shows up as a strong power-law increase of both, $\varepsilon'$ and $\varepsilon''$ at low frequencies, which is most pronounced for higher temperatures. It leads to extremely large values of $\varepsilon'$ at low frequencies (not completely shown in Fig. 2), typical for non-intrinsic effects caused by electrode polarization.[67] For the present samples, it most likely is caused by the unavoidable ionic charge transport within hydrous samples, which essentially becomes blocked at the electrodes, grain boundaries, and/or pore walls. The relaxational nature of process 2 is clearly revealed by the step in $\varepsilon'(\nu)$ [e.g., at about 10$^4$ Hz for the 210 K curve; Fig. 2(a)]. Due to the overlap with process 1, the corresponding loss peak is only visible as a shoulder [Fig. 2(b)]. In contrast, the relaxational process 3 is well pronounced both in $\varepsilon'(\nu)$ and $\varepsilon''(\nu)$. Finally, in the loss spectra of Fig. 2(b), at low temperatures another fast process is evidenced by an excess contribution showing up at high-frequencies at the right flanks of the peaks associated with relaxation 3. Its loss peaks are only incompletely seen in the covered frequency range and the corresponding $\varepsilon'(\nu)$ step is too weak to be detectable.

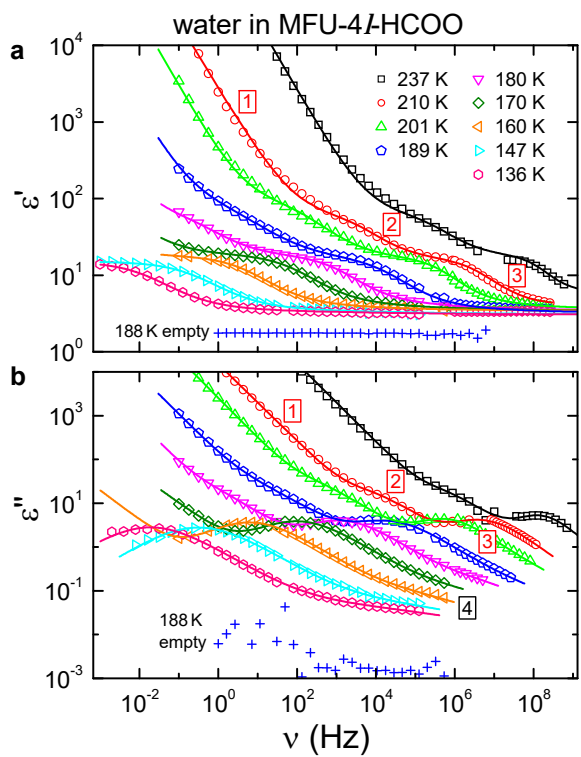

**Fig. 2 Dielectric spectra of confined water.** The dielectric constant (a) and loss (b) of water confined in MFU-4*l*-HCOO are shown for selected temperatures. The lines are simultaneous fits of the real and imaginary parts employing Cole-Cole functions for the three observed relaxation processes and a power law for a phenomenological description of the strong low-frequency increase (for details see text). The numbers indicate the different processes (1 - 3 are indicated for the 210 K spectrum; 4 denotes the fastest process, detected at low temperatures only). For comparison, the plusses show results for the pure MOF without water ("empty") at 188 K.

Process 2 clearly is much slower than the relaxation dynamics of supercooled water that is unaffected by pore-wall interactions. This can be concluded from a comparison with results on supercooled bulk water (however, only available at temperatures beyond the borders of the no-man's land), aqueous water solutions, and water confined in different host systems as MCM-41.[19,45,68] It may be ascribed to water molecules interacting with the pore walls (probably by forming hydrogen bonds with the HCOO groups), which considerably slows down the molecular motions as often found in confined systems.[1,2,8,36,37,69,70,71] However, other explanations also are possible, e.g., by a partial solvation of the formate ion or in terms of a non-intrinsic Maxwell-Wagner relaxation, arising from the heterogeneous nature of the guest-host system.[39] In any case, this process does not reflect any dynamics close to that of bulk water[19,45] and therefore will be not be treated here in further detail.

In contrast, the characteristic time scale of process 3, estimated by its loss-peak frequency [Fig. 2(b)], lies well within the region often considered for the "true" water dynamics of supercooled water within the no-man's land.[19,45] For example, the peak frequencies are located at similar frequencies as for the "ideal" confinement host[19,45] MCM-41 with 2.1 nm pore size (cf., e.g., Fig. 3 of Ref. 39). As both relaxational features, the $\varepsilon'$ step and $\varepsilon''$ peak, are well resolved in the spectra, reliable information on the temperature-dependence of its relaxation time can be obtained by fitting the data as described below.

Relaxation 4 obviously is significantly faster than the time scale of the bulk-water dynamics supposed for the no-man's land.[19,45] Various explanations of this process are possible. For example, it may be ascribed to water within the smaller pores of about 1.2 nm diameter existing in MFU-4*l*-HCOO. There the stronger confinement could lead to a considerable acceleration of molecular dynamics because the development of intermolecular cooperativity (which slows down the dynamics) can be suppressed in small pores. Such behavior is often found in supercooled liquids and can be used to estimate the size of the cooperativity length scale in dependence of temperature (see, e.g., Refs. 1,11,47). Within the framework of a cooperativity-driven glass transition,[72,73,74,75] this would imply that, for the temperature range where this process is detected (about 136 - 180 K), the cooperativity length in bulk water always exceeds the pore size of about 1.2 nm. As an alternative, process 4 can be interpreted as a secondary relaxation associated to the main water relaxation given by process 3. The occurrence of secondary relaxations, also termed $\beta$ relaxations, which are faster than the $\alpha$ relaxation that reflects the main molecular motion governing the glass transition, is a common finding for supercooled liquids.[76,77,78] To distinguish such secondary relaxations from those sometimes arising from molecule-specific degrees of freedom like side-chain motions in polymers, they are commonly termed Johari-Goldstein (JG) relaxation.[76] Indeed, very recently a rather fast water relaxation (with relaxation times around $10^{-5}$ s at 140 K), found in various confinements, was proposed to be of JG type.[46] However, one should be aware that in several works the much slower main confined-water relaxation, corresponding to process 3 of the present work, was also suggested to reflect the JG relaxation of bulk water.[19,45,79] In Ref. 46 this apparent discrepancy was resolved by proposing two different $\beta$ relaxations for confined water. It is clear that, based on the present data that do not fully resolve this process, we cannot clarify its true origin. Especially, it is not possible to deduce reliable values for the relaxation times from the fits of the spectra, described below. While the present work does not aim at the clarification of the origin of this process, nevertheless it is worth to mention that the low-temperature loss spectra of Fig. 2(b) closely resemble those of supercooled liquids with a $\beta$ relaxation that is submerged under the dominating $\alpha$ peak and leads to a so-called excess wing.[80,81] This also includes aqueous solutions like $H_2O$:LiCl.[82]

**Relaxation time.** Of the detected processes, obviously relaxation 3 is the most relevant one to be compared to water dynamics in pure water and in other confinement systems as MCM-41. To determine its most important parameter, the relaxation time $\tau$, we have fitted our broadband spectra (for a discussion of the other parameters of relaxation 3, see Supplementary Information). To account for process 1 in the fits, we phenomenologically described it by power laws $\varepsilon'' = a\, \nu^{s-1}$ and $\varepsilon' = a \tan(s\pi/2)\, \nu^{s-1}$ with an exponent parameter $s < 1$ and $a$ denoting a prefactor (the additional tangent prefactor for $\varepsilon'$ ensures compliance with the Kramers-Kronig relation[83]). The three relaxational processes were described by a sum of Cole-Cole functions, an empirical formula often used to parameterize dielectric relaxation processes.[84,85] One should be aware that, depending on temperature, only part of these contributions had to be included in the overall fit function, limiting the total number of parameters. For example, at high temperatures, process 4 and at low temperatures, process 1 and 2 have shifted out of the frequency window. The lines in Fig. 2 are fits, simultaneously performed for the real and imaginary part of the permittivity. Obviously, a reasonable agreement of fits and experimental spectra is achieved in this way.

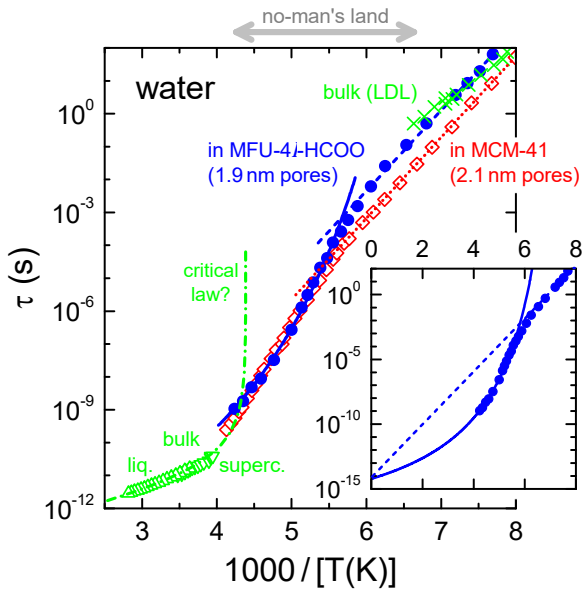

**Fig. 3 Arrhenius plot of the relaxation times of confined and bulk water.** Confined-water data in MFU-4*l*-HCOO for process 3 are shown by the closed circles. For comparison, the diamonds show data for water in MCM-41 with 2.1 nm pore size,[39] taken from Ref. 19. In addition, relaxation times for liquid[82] and supercooled bulk water[68] are included at high temperatures (upright and inverted triangles, respectively) and for bulk LDL water at low temperatures (crosses).[86] The dash-dotted line is a critical law with $T_c$ = 228 K, proposed for bulk water.[46,106] The solid line is a fit of the high-temperature ($T > 175$ K, i.e., $1000/T < 5.7$ K$^{-1}$) data of process 3 with the VFT function (eq. 1; parameters: $T_{VF}$ = 122 K, $D$ = 11.3, $\tau_0$ = 6.2×10$^{-15}$ s). The dashed line at lower temperatures represents an Arrhenius law with its energy barrier of 0.48 eV fixed to the same value as for the low-temperature data of water in MCM-41 (dotted line). The approximate extent of the non-man's land is indicated by the double-arrow on top of the figure. The inset demonstrates that $\tau(T)$ of process 3 can be described by the VFT (above $T_{cr}$) and Arrhenius laws (below $T_{cr}$) using identical prefactors $\tau_0$, which can be read off at $1000/T = 0$.

Figure 3 shows an Arrhenius representation of the temperature-dependent relaxation times of process 3 as obtained from these fits (closed circles). For comparison, data on MCM-41 with 2.1 nm pore size[19,39] are included in the figure (diamonds) as well as results on bulk water at high temperatures[68,82] (triangles) and on bulk low-density water well below the no-man's land (crosses).[86] At high temperatures, both $\tau(1/T)$ traces come close to that of liquid and supercooled bulk water, shown by the triangles.[68,82] This supports the notion[19,45] that water in MCM-41 represents a reasonable model system of confined water, which obviously also is the case for hydrous MFU-4*l*-HCOO. A closer look, however, reveals that both data sets do not seem to provide a perfect low-temperature continuation of the high-temperature bulk results measured outside the no-man's land. This is consistent with (highly disputed)[28,29,30,31,32,33,34] ideas that the true bulk-water data would exhibit a very strong upturn just after entering the no-man's land under (super)cooling (dash-dotted line in Fig. 3)[106] and that the MCM-41-confined water dynamics essentially reflects the $\beta$ relaxation of water.[19,46]

While the $\tau(1/T)$ data for the MFU-4*l*-HCOO host at $T > 175$ K ($1000/T < 5.7$ K$^{-1}$) are of similar order of magnitude as the corresponding data in MCM-41, they exhibit a remarkable difference: In contrast to MCM-41, they reveal clear deviations from simple thermally activated behavior, $\tau = \tau_0 \exp[E/(k_B T)]$ (where $E$ is an energy barrier and $\tau_0$ an inverse attempt frequency). Instead they can be reasonably described by the Vogel-Fulcher-Tammann (VFT) law (solid line in Fig. 3), written in its modified form as proposed by Angell:[87]

$$\tau = \tau_0 \exp\left[\frac{DT_{VF}}{T - T_{VF}}\right] \qquad (1)$$

Here $T_{VF}$ denotes the Vogel-Fulcher temperature, where $\tau$ diverges and the strength parameter $D$ characterizes the deviation from Arrhenius behavior.[87] VFT behavior is commonly found for the $\alpha$ relaxation of supercooled liquids and regarded as typical for glassy freezing.[81,85,87,88] It can be ascribed to increasingly cooperative molecular motions when approaching the glass transition.[72,73,74,75,89] Non-Arrhenius behavior of confined water at temperatures above about 180 K, based on dielectric spectroscopy, was also discussed by some earlier works.[19,39,44,79] However, either at high temperatures only Arrhenius behavior was detected in these works, or the VFT temperature dependence was based on rather few data point only and, to our knowledge never, revealed as clearly as in Fig. 3. The effective 3D confinement, reduced pore-wall interactions, the performed evaluation of real and imaginary part including all processes and/or the extension of our spectra up to 1 GHz, leading to better precision of the high-temperature values of $\tau$, may explain this fact. In any case, the steeper high-temperature Arrhenius region, revealed by hydrous MCM-41 (diamonds in Fig. 3 at $1000/T < 5.7$ K$^{-1}$),[19,39] leads to an unphysical attempt frequency, $1/(2\pi\tau_0)$, which should be of the order of a typical phonon frequency but is many decades smaller (about $10^{-27}$ s).[39] This indicates that deviations from the Arrhenius law must also be present for this host material.[39] The well-defined VFT law for water in MFU-4*l*-HCOO allows to determine its glass temperature using the usual definition $\tau(T_g) \approx 100$ s, which leads to $T_g \approx 159$ K. Its strength parameter $D \approx 11.3$, corresponding to a fragility parameter of $m \approx 68$,[90] characterizes this water phase in confinement as moderately fragile within the strong/fragile classification scheme proposed by Angell.[87] The notion of a liquidlike non-Arrhenius behavior of water confined in MCM-41 is consistent with Ref. 91, where the authors point out that the found specific-heat anomaly at roughly 225 K, observed in MCM-41 with 1.8 nm pores, is not due to a glass transition.

Notably, the $\tau(1/T)$ curves of both water in MCM-41 and in MFU-4$l$-HCOO exhibit a change from stronger to weaker temperature dependence below a crossover temperature $T_{cr} \approx 175$ K ($1000/T > 5.7$ K$^{-1}$; Fig. 3). In both cases, at $T < T_{cr}$ Arrhenius behavior is found with essentially identical activation energies of $\approx 0.48$ eV, as evidenced by the similar slope of their $\tau(1/T)$ curves in Fig. 3. In this region, however, water in MFU-4$l$-HCOO relaxes by about a factor of 10 slower than in MCM-41. Interestingly, at low temperatures our data rather well match those reported for the low-density liquid (LDL) phase of bulk water (also termed low-density amorphous ice), which were measured beyond the low-temperature border of the non-man's land.[86]

## Discussion

For the change in the temperature dependence of $\tau(T)$, as observed in MCM-41 at about 180 K and as also found in several other confinement systems,[18,19,44] various explanations were proposed. This includes the interaction of the water molecules with the pore walls[45] or an artifact of the analysis[38] that can arise when the $\beta$ and $\alpha$ relaxations merge.[92] Moreover, various explanations in terms of a $\beta$ relaxation were discussed, either assuming a transition from $\alpha$-like to $\beta$ dynamics[18,19,44,46] or in terms of a change of the temperature dependence of the $\beta$ relaxation alone.[17,93] The latter would correspond to the crossover of the $\beta$-relaxation time $\tau_\beta(T)$ from Arrhenius to VFT behavior, as often seen in bulk supercooled liquids when crossing $T_g$ under heating.[77,88,94,95,96,97] This scenario thus would imply a glass transition of $T_g = T_{cr} \approx 175$ K for water confined in MFU-4$l$-HCOO.

However, an alternative explanation of our $\tau(T)$ results arises when considering their qualitative similarity to findings for confined glassforming liquids.[1,11,47,56] The motivation of such confinement studies originally was the verification of increasing cooperativity length scales $\xi$ arising when approaching $T_g$. As mentioned above, such a scenario was suggested for the explanation of the common non-Arrhenius behavior of the $\alpha$ dynamics of bulk systems, assuming that the glass transition is due to an underlying phase transition.[72,73,74,75] Indeed, when overcoming obstacles as pore-wall interactions or pore sizes that are much smaller or larger than $\xi$, a dynamical crossover of the $\alpha$ relaxation time from bulklike VFT temperature dependence to Arrhenius behavior could be observed under cooling for various glassforming liquids.[1,11,47,56] It occurs when $\xi$ exceeds the pore diameter and, thus, cannot grow any further. Using hosts with different pore diameters then can provide information about the temperature-dependent cooperativity length.[11]

Therefore it seems natural to explain our results in confined water in the same way: Process 3 is the $\alpha$ relaxation of confined supercooled water (however, differing significantly from bulk water as discussed below) with $T_g \approx 159$ K. With decreasing temperature, at $T_{cr} \approx 175$ K its cooperativity length has grown to about 1.9 nm and, without confinement, would exceed the pore size under further cooling. Below $T_{cr}$, the detected relaxation still reflects cooperative $\alpha$ dynamics, but with temperature-independent cooperativity because $\xi$ is limited by the pore diameter $d$. This leads to the observed Arrhenius behavior, in agreement with model considerations[98] based on the Adam-Gibbs theory.[72]

Interestingly, within this scenario the approximate agreement of $\tau(1/T)$ at low temperatures with the relaxation times reported for bulk LDL water (Fig. 3) are consistent with ideas that the latter do not reflect the $\alpha$ dynamics of supercooled bulk water with freely unfolded cooperativity. In Ref. 19 it was suggested that the methods used to produce LDL water[86,99] may not be able to create the same type of water as could be reached by slow supercooling if crystallization would not intervene. For hyperquenched LDL water, this is supported by theoretical computations revealing characteristic length scales of less than about 3 nm, consistent with the present results.[100] In some respect, the detected confined water dynamics at $T < T_{cr}$ reminds of a $\beta$ relaxation, which is often assumed to have no or only minor cooperativity. This view is, e.g., advocated in Ref. 45, based on an estimate of the $\beta$-relaxation time of bulk water using the coupling model,[101,102] which leads to values consistent with those experimentally found for hydrous MCM-41 at low temperatures, $T < T_{cr}$.[45]

The above rationalization of the crossover at $T_{cr}$ is supported by the fact that both the VFT and the Arrhenius regions of the $\tau(1/T)$ curve can be consistently described assuming the same attempt frequency, $1/(2\pi\tau_0)$, as demonstrated in the inset of Fig. 3. Further support for this cooperativity-related explanation is provided by Ref. 44 where such a scenario for confined water was already previously considered. There a systematic variation of the crossover temperatures in dependence of the pore sizes of various confinement hosts was found, just as expected: Under cooling, for larger pores $\tau(1/T)$ should adhere to VFT behavior down to lower temperatures, because $\xi$ can grow further before exceeding the pore diameter at $T_{cr}$. One should be aware that the absolute values of $\xi$ deduced from such studies can only be estimates. For example, NMR experiments on water in MCM-41 have revealed the formation of crystalline water fractions near the pore center[17,93] and pore-size dependent partial crystallization was also considered in other works.[103,104,105] This implies a reduced confinement volume. In addition, a layer of slowed-down molecules interacting with the pore walls, which should depend on the used host material, may also reduce the effective pore size. Interestingly, $T_{cr}$ is somewhat higher for MCM-41 than for MFU-4$l$-HCOO (see the crossing points of the solid and dashed lines in Fig. 3) despite the nominally larger pore size of MCM-41 (2.1 instead of 1.9 nm). This is consistent with the suggested stronger pore-wall interactions in MCM-41 and the partial water crystallization in its pores, which both should reduce the effective pore volume available for cooperative motions.

One should note that, most likely, the confined $\alpha$ dynamics at $T > T_{cr}$ does not reflect that of the bulk if it could be detected within the no man's land. This is evidenced by the mentioned one-decade offset of the (extrapolated) $\tau(1/T)$ traces of bulk and confined water showing up at about 250 K (Fig. 3). To explain this finding, one may assume that the $\alpha$-relaxation time of bulk water exhibits an extreme upturn of $\tau(1/T)$, just after entering the no-man's land, as shown by the dash-dotted line in Fig. 3. The latter represents a critical law with $T_c = 228$ K as proposed in Ref. 106. Such a scenario was considered, e.g., in Refs. 46 and 79, where the confined water dynamics, even at $T > T_{cr}$, consequently was denoted as $\beta$ relaxation. As discussed in these works, indeed in some respect the confined dynamics detected below $T_c$ has $\beta$-like characteristics. However, one should be aware that, within this scenario, the glass temperature of bulk water [usually defined by $\tau(T_g) \approx 100$ s] would be only slightly above $T_c = 228$ K (this becomes obvious when extrapolating the critical law in Fig. 3 to 100 s). This would imply that nearly all the confinement data of Fig. 3 were collected below $T_g$. However, as mentioned above, below the glass transition, $\tau(T)$ of a $\beta$ relaxation should exhibit Arrhenius behavior instead of the VFT temperature dependence observed in the present work. Even the $\alpha$-relaxation time only follows VFT temperature dependence at $T > T_g$,[88,107,108] except for extremely slow cooling or if $\tau$ is deduced from aging experiments.[109] Thus, even within the scenario of a critical water temperature of 228 K, we suggest that the observed relaxation process should be denoted as the $\alpha$ relaxation of confined water. It is this cooperative molecular motion that governs the glassy freezing of this special supercooled water phase in confinement and there is no slower relaxation that could assume the role of the $\alpha$ relaxation in confinement.

Our results imply that the MOF MFU-4l-HCOO represents an ideal confinement host with superior properties as effective 3D pore geometry and highly hydrophobic pore walls. Water confined in this system in several respects behaves similar as in MCM-41, previously considered as ideal confinement system.[19,45] However, in marked contrast to the latter and much clearer than in any other host system, our measurements up to 1 GHz reveal significant deviations from simple thermally activated dynamics above a crossover temperature $T_{cr} \approx 175$ K, where $\tau(T)$ follows the VFT law. This evidences cooperative, $\alpha$-like relaxation dynamics of confined water, whose cooperativity length increases under cooling in this temperature region. Below $T_{cr}$, $\tau(T)$ still reflects cooperative dynamics but with temperature-independent cooperativity length $\xi$, limited by the pore size of 1.9 nm as also found in supercooled liquids like glycerol.[1,11,47,56] As discussed above, the observed confined-water dynamics, both above and below $T_{cr}$, in some respects exhibits the characteristics of a secondary relaxation as pointed out in several previous confinement studies of water (e.g., 19,45,46). However, it can be considered as the $\alpha$ process of confined water (in accord with the term $\alpha_{conf}$ in Ref. 19) as it governs the cooperativity-driven glassy freezing of this water phase.

As revealed by Fig. 3, the deduced $\tau(T)$ data of water confined in MFU-4l-HCOO at first glance provide a reasonable continuation of the high-temperature bulk data into the no-man's land (however, with some offset) and closely match the relaxation times of bulk LDL water at low temperatures. Thus, it is tempting to simply claim that the present results provide information on the properties of hypothetical supercooled bulk water within the no-man's land, which would imply a liquid-liquid transition at 175 K. However, in fact the situation is not so straightforward and the detected confined-water relaxation represents the dynamics of a different kind of water, whose behavior is significantly modified due to the confinement. Interestingly, at least at $T > T_{cr}$ this modification does not arise from the usual restriction of the molecular cooperativity associated with the glassy freezing as found in other confined liquids. This is signified by the detected non-Arrhenius behavior evidencing unimpeded growth of $\xi(T)$ under cooling. These findings thus point to the presence of two different length scales in water: The smaller one, also found in other supercooled liquids, is the cooperativity length scale $\xi$ ($< 1.9$ nm for $T > T_{cr}$) governing glassy freezing of confined water in terms of a hidden "amorphous-order" phase transition underlying the glass transition of all supercooled liquids.[73,75,89] The longer one (significantly larger than 1.9 nm) represents the characteristic length scale of an extended network of hydrogen-bonded molecules, which is essential for the specific molecular dynamics of bulk water. This length scale may, e.g., be related to the "patches" of four-bonded molecules considered within the percolation theory of water by Stanley and Teixeira.[110] As such an extended network cannot form in nm-sized pores, markedly different but still cooperative relaxation dynamics is revealed for water in confinement. The glass temperature of the resulting confined water phase is 159 K, it behaves moderately fragile ($m \approx 68$), and, for infinitely slow cooling rate, its cooperativity length should diverge close to the Vogel-Fulcher temperature $T_{VF} = 122$ K. Similar considerations involving such a network were also made in Refs. 15 and 46 to explain the absence of a bulklike $\alpha$ relaxation in confined water which now is confirmed by the experimental results of the present work. In Ref. 91 it also was suggested that the restricted pore size affects the hydrogen bonding and the resulting arrangement of the water molecules, leading to properties that differ from those of bulk water. There the confined-water state in MCM-41 was described as distorted ice-like structure in equilibrium with the water melt. As shown in the present work, water in MFU-4l-HCOO at least exhibits significant liquidlike fractions with relaxation times considerably faster than in ice.[111,112]

To corroborate the scenario promoted in the previous paragraph, measurements in MOFs with different pore diameters should be performed. The use of pores that are sufficiently large to allow for the unfolding of the extended hydrogen-bonded network, thus leading to bulklike properties, may be limited by the increasing importance of crystallization.[103,105] However, pore sizes only somewhat smaller and larger than 2 nm should be able to reveal the temperature dependence of the cooperativity length $\xi$ governing the glass transition of confined water.

## Methods

**Host material synthesis.** MFU-4*l* was synthesized according to the procedure described in Ref. 63. MFU-4*l*-HCOO, used as confinement host in the present work, was prepared by postsynthetically exchanging the chloride side-ligands in MFU-4 with HCOO-anions, similar to the procedure described in Ref. 65. The synthesis and structural parameters of MFU-4*l*-HCOO have been reported in the Supporting Information file of Ref. 66.

**Water content calculation.** The theoretical water content of MFU-4*l*-HCOO was calculated as follows: In a volume of 1000 Å$^3$, a content of 34 $H_2O$ molecules leads to a density of ca. 1.0 g/cm$^3$. The void volume of MFU-4*l* is about 23575 Å$^3$ (ca. 78.7%) which would correspond to about 800 $H_2O$ molecules per unit cell. For the structural models, the initial positions of the water molecules were obtained from sorption calculations (Sorption Module in the Materials Studio 2019 software) employing force-field (Dreiding) derived energy values and partial charges (Qeq). In a first run, an empty box was filled with $H_2O$ molecules and the final fugacity was varied until a water density of about 1.0 g/cm$^3$ was reached. The same fugacity parameter (20000 kPa) was then employed to calculate the position and numbers of water molecules in MFU-4*l*-HCOO. (NB: The very high fugacity required for this is due to the Dreiding force-field parameter sets who have not been optimized for condensed phases). These simulations led to a final value of 720 $H_2O$ molecules per unit cell.

**Sample preparation.** The host material was first degassed for 24 h at 150 °C in vacuum. The loading with water was directly done within the dielectric measurement setup by exposing the degassed samples to a water-vapor atmosphere at room temperature. After 24 h the permittivity stopped increasing, indicating full saturation of the pores, whereupon the temperature-dependent dielectric measurements were started. The permittivity is a well-suited quantity to monitor the water content because water has a high dipolar moment and also introduces considerable ionic conduction, thus leading to strong contributions in both the real and imaginary part of the permittivity.

**Dielectric measurements.** For the dielectric measurements, the powder samples were filled into parallel-plate capacitors with a plate distance of 70 μm. Using powder avoids possible pressure-induced deterioration of the host material that could arise when pressing pellets. Only slight pressure was applied to the capacitor plates to enhance the packing density. Nevertheless, the detected absolute values of the dielectric permittivity may be somewhat reduced due to a packing density less than one. The dielectric measurements of the complex permittivity at frequencies from 1 mHz up to 1 GHz were performed using two experimental techniques:[113] At frequencies up to about 3 MHz, a frequency-response analyzer (Novocontrol Alpha) was employed. High-frequency measurements up to about 1 GHz were carried out using a coaxial reflectometric setup[114] with an impedance analyzer (Keysight E4991A). For cooling, a closed-cycle refrigerator was used.

**Data availability**
The data that support the findings of this study are available from the corresponding author upon reasonable request.

**Acknowledgements**
D.V. is grateful to the DFG for financial support (DFG priority programme 1928 Coordination Networks: Building Blocks for Functional Systems, "COORNETs").

**Author contributions**
A.L., P.L. and D.V. conceived and supervised the project. J.K.H.F. and P.S. performed the dielectric measurement. The results were analyzed by J.K.H.F. and P.L. D.D. prepared the host systems. D.V. performed molecular simulations and determined the theoretical filling grade. P.L. wrote the paper.

**Competing interests**
The authors declare no competing interests.

**Additional information**
Correspondence and requests for materials should be addressed to P.L.

**Supplementary Figures**

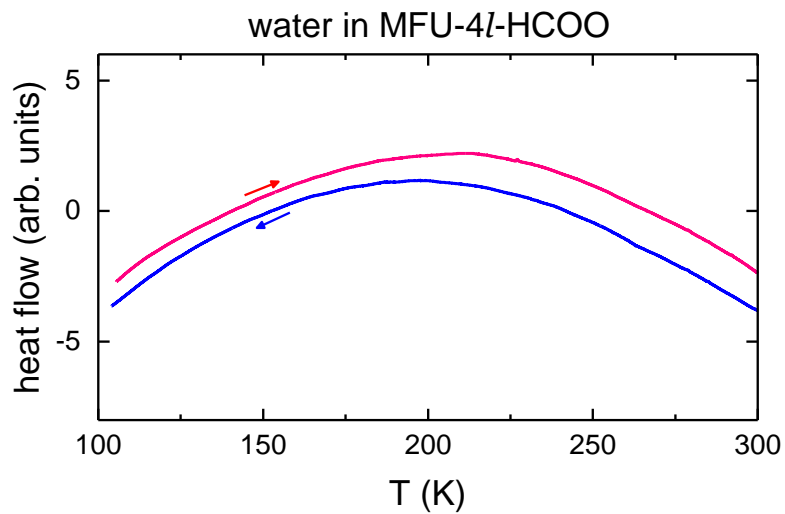

**Supplementary Figure 1: DSC heat flow for water-filled MFU-4*l*-HCOO.** Data are shown for cooling and heating as indicated by the arrows. The broad peak-shaped behavior arises from the instrument background.

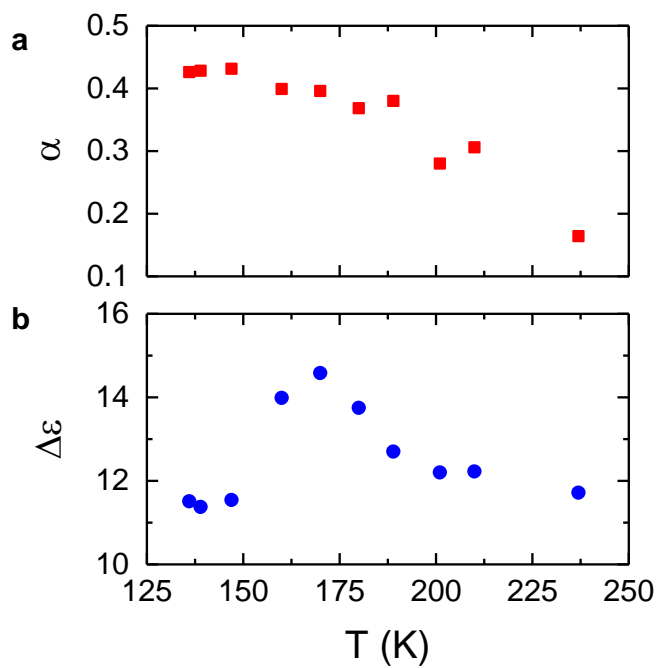

**Supplementary Figure 2: Parameters of process 3.** The figure shows the temperature dependence of the width parameter (a) and the relaxation strength (b) for relaxation 2. The data were obtained from fits as shown in Fig. 2.

**Supplementary Discussion**

Supplementary Figure 1 shows the DSC heat flow for water-filled MFU-4l-HCOO for cooling and heating. The broad peak-shaped behavior arises from the instrument background. There are no indications of anomalies and, thus, of water crystallization in the investigated temperature range.

Supplementary Figure 2 presents the width parameter $\alpha$ (a) and the relaxation strength $\Delta\varepsilon$ (b) for relaxation 3, the $\alpha$-relaxation of water confined in MFU-4l-HCOO. These data were obtained from fits as described in the main text and shown in Fig. 2. For the fits, relaxation 3 was described by the Cole-Cole equation:[1]

$$\varepsilon^*(\nu) = \varepsilon_\infty + \frac{\Delta\varepsilon}{1 + (i2\pi\nu\tau)^{1-\alpha}}$$

Here $\varepsilon^* = \varepsilon' - i\varepsilon''$ is the complex permittivity, $\varepsilon_\infty$ the high-frequency limit of the dielectric constant, $\Delta\varepsilon$ is the relaxation strength, and $\tau$ the relaxation time. The parameter $\alpha \leq 1$ is a width parameter leading to a symmetric broadening of the loss peaks in relation to the Debye function, which is recovered for $\alpha = 0$. Deviations from the Debye case usually signal a distribution of relaxation times, typical for disordered systems like glassforming liquids.[2] Supplementary Figure 2(a) reveals a systematic decrease of $\alpha(T)$ with increasing temperature, especially at high temperatures, implying a successive narrowing of the loss peak width. Such a behavior is rather common for the $\alpha$ relaxation of glassforming liquids and may be ascribed to increasingly faster thermal fluctuations at high temperatures, which blur the heterogeneities causing the broadening.[3]

The relaxation strength shown in Supplementary Figure 2(b) varies relatively weakly and exhibits a peak close to the crossover temperature $T_{cr} \approx 175$ K of the relaxation time (cf. Fig. 3). Below $T_{cr}$, $\tau(T)$ crosses over to a weaker temperature dependence, which is interpreted as a confinement effect as discussed in detail in the main text. The increase of $\Delta\varepsilon(T)$ with decreasing temperature, observed above about 170 K, corresponds to the canonical behavior of dipolar systems as expected, e.g., within the time-honored Onsager theory.[4] The found non-canonical peak in $\Delta\varepsilon(T)$ close to $T_{cr}$, at first glance seems to be connected to the confinement-induced suppression of the increase of the cooperativity length under cooling, discussed in the main text. However, one should be aware that unconfined LiCl-water solutions, which can be easily supercooled, seem to exhibit a similar peak in $\Delta\varepsilon(T)$.[5,6] Thus, currently it is not clear whether the peak, revealed in Supplementary Figure 2(b), is due to the confinement or a genuine bulk property of supercooled water. This interesting question should be clarified by further investigations, e.g., by performing broadband dielectric measurements for other aqueous solutions.

**Supplementary References**